\documentclass[aps,prl,reprint,amsmath,amssymb,superscriptaddress]{revtex4-2}
\usepackage{amsmath} 
\usepackage{graphicx}
\usepackage{dcolumn}
\usepackage{bm}
\usepackage{physics}
\usepackage{color}
\usepackage{booktabs}
\usepackage[colorlinks=true, citecolor=blue, linkcolor=blue, urlcolor=blue]{hyperref}
\usepackage{color}
\usepackage{graphicx}
\usepackage{algpseudocode}

\usepackage{changes}

\begin{document}

\preprint{APS/123-QED}
\title{Respective Roles of Electron-Phonon and Electron-Electron Interactions\\ 
in the Transport and Quasiparticle Properties of SrVO$_3$}

\author{David J. Abramovitch}%
\affiliation{Department of Applied Physics and Materials Science, and Department of Physics, California Institute of Technology, Pasadena, California 91125}
\affiliation{Center for Computational Quantum Physics, Flatiron Institute, 162 5th Avenue, New York, New York 10010, USA}
\author{Jernej Mravlje}
\affiliation{Jožef Stefan Institute, Jamova 39, 1000 Ljubljana, Slovenia}
\affiliation{Faculty of Mathematics and Physics, University of Ljubljana, Slovenia}
\author{Jin-Jian Zhou}
\affiliation{School of Physics, Beijing Institute of Technology, Beijing, 100081, China}
\author{Antoine Georges}
\affiliation{Coll\`ege de France, Paris, France and CCQ-Flatiron Institute, New York, NY, USA}
\affiliation{Center for Computational Quantum Physics, Flatiron Institute, 162 5th Avenue, New York, New York 10010, USA}
\affiliation{Centre de Physique Théorique, Ecole Polytechnique, CNRS, Institut Polytechnique de Paris, 91128 Palaiseau Cedex, France}
\affiliation{DQMP, Université de Genève, 24 quai Ernest Ansermet, CH-1211 Genève, Suisse}
\author{Marco Bernardi}
\email{bmarco@caltech.edu}
\affiliation{Department of Applied Physics and Materials Science, and Department of Physics, California Institute of Technology, Pasadena, California 91125}


\begin{abstract}
The spectral and transport properties of strongly correlated metals, such as SrVO$_3$ (SVO), 
are widely attributed to electron-electron ($e$-$e$) interactions, with lattice vibrations (phonons) playing a secondary role.
Here, using first-principles electron-phonon ($e$-ph) and dynamical mean field theory calculations, we show that $e$-ph interactions play an essential role in SVO: they govern the electron scattering and resistivity in a wide temperature range down to 30 K, and induce an experimentally observed kink in the spectral function. 
In contrast, the $e$-$e$ interactions control quasiparticle renormalizations and low temperature transport, 
and enhance the $e$-ph coupling.
We clarify the origin of the near $T^2$ temperature dependence of the resistivity by analyzing the $e$-$e$ and $e$-ph limited transport regimes. Our work disentangles the electronic and lattice degrees of freedom in a prototypical correlated metal, revealing the dominant role of $e$-ph interactions in SVO.
\end{abstract}
\maketitle
%

\begin{figure}[t!]
    \centering
    \includegraphics[width = 0.5\textwidth]{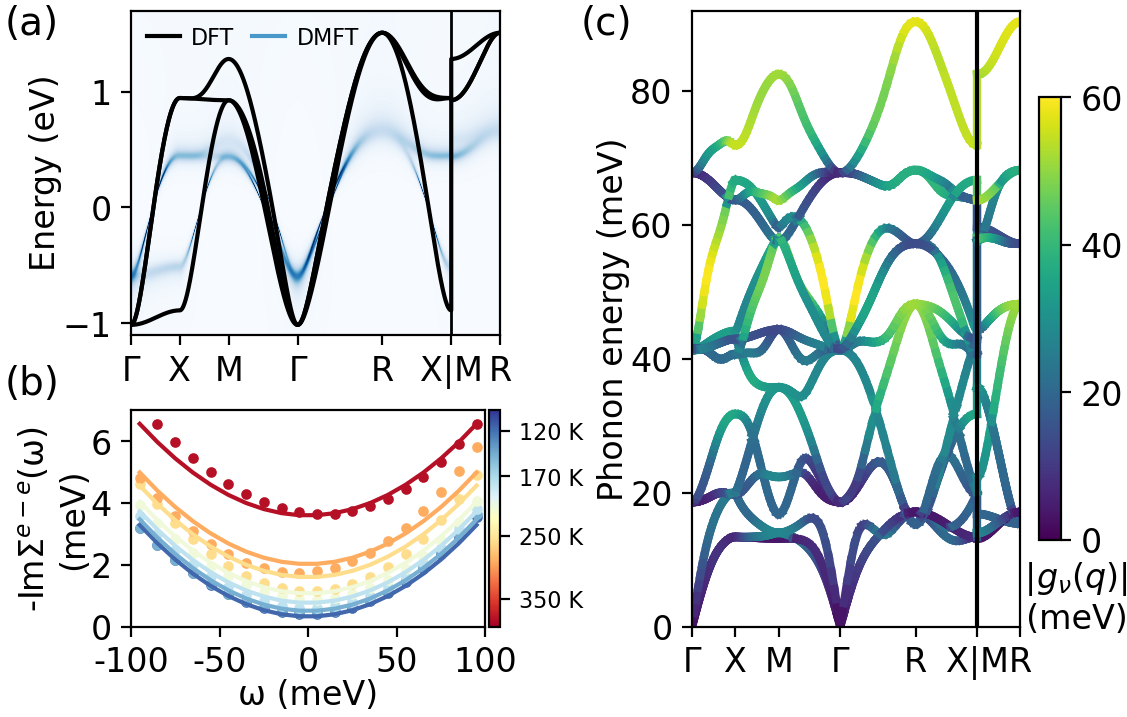}
    \caption{(a) DFT electronic band structure (black) and the spectral function computed with DMFT at 290 K (blue), 
    showing renormalization by a factor $Z \approx 0.5$. (b) Imaginary part of the electron self-energy due to $e$-$e$ interactions, computed with DMFT. The lines show a fit to the Fermi liquid form, $\mathrm{Im}\Sigma^{\rm e-e}(\omega,T) = -c((\hbar\omega)^2 + \pi^2 (k_B T)^2)$ with $c \approx 0.33 \mathrm{eV}^{-1}$. (c) DFPT phonon dispersions in SVO, with colors showing the $e$-ph coupling strength $|g_{\nu}(\mathbf{q})|$ averaged on the Fermi surface. }
    \label{fig:bands-spec-imsigma}
\end{figure}

\begin{figure}[!hb]
    \centering
    \includegraphics[width = 0.5\textwidth]{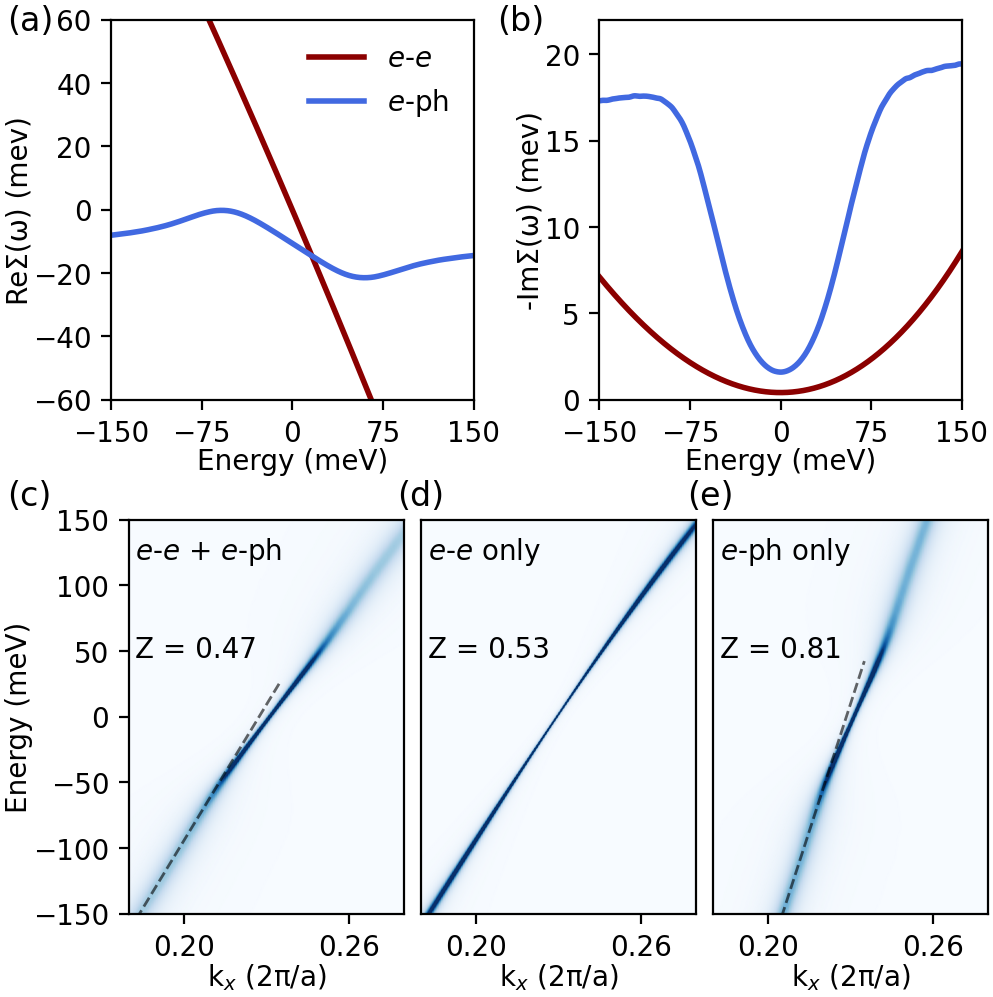}
    \caption{(a) Real and (b) imaginary parts of the self-energy at $T=115$K and $\mathbf{k}$ = (0.23$\times 2\pi/a$,0,0), showing contributions from $e$-$e$ and $e$-ph interactions. (c) Spectral functions including both $e$-$e$ and $e$-ph interactions, (d) $e$-$e$ interactions only, and (e) $e$-ph interactions only. Quasiparticle weights $Z$ are indicated for each spectral function, and the dashed lines in (c) guide the eye to the quasiparticle dispersion near the kink. All spectral functions are shown along the $\Gamma$$-$X direction near the Fermi surface at 115 K. 
    }
    \label{fig:se-spec}
\end{figure}

\textit{Introduction.\textemdash}
Strontium vanadate, SrVO$_3$ (SVO), is a perovskite \mbox{oxide} widely studied as a prototypical correlated metal~\cite{onoda_metallic_1991,lan_structure_2003, zhang_correlated_2016}. Experiments have measured transport and spectral functions in detail in SVO, owing to advances in growth of clean samples~\cite{brahlek_accessing_2015, brahlek_mapping_2016} and characterization by angle-resolved photoemission spectroscopy~\cite{yoshida_direct_2005,kobayashi_origin_2015, yoshida_correlated_2016}. 
There are clear spectroscopic signatures of strong electron interactions in SVO, including kinks in the quasiparticle dispersion~\cite{aizaki_self-energy_2012, yoshida_correlated_2016} and mass enhancement with quasiparticle weight~$Z \!\approx \!0.5$~\cite{yoshida_direct_2005}. 
In addition, transport measurements have found a near $T^2$-dependent resistivity in broad temperature ranges below 300~K~\cite{moyer_highly_2013,fouchet_study_2016, mirjolet_electronphonon_2021, ahn_low-energy_2022, brahlek_hidden_2024}. 
\\
\indent
These findings are often attributed to strong electron-electron ($e$-$e$) interactions. 
As a result, SVO serves as a testbed for theoretical methods treating strongly correlated materials, including first-principles variants of dynamical mean field theory (DMFT) such as density functional theory (DFT)+DMFT~\cite{nekrasov_momentum-resolved_2006}, GW+DMFT~\cite{tomczak_combined_2012, taranto_comparing_2013,sakuma_electronic_2013, tomczak_asymmetry_2014}, and linear response DMFT~\cite{kocer_efficient_2020}, and the dynamical cluster~\cite{lee_dynamical_2012} and dynamical vertex approximations~\cite{galler_ab_2017}. 
\\
\indent
However, one can question whether the transport properties and spectral features observed in SVO are the result of purely electronic interactions. 
In particular, electron-phonon ($e$-ph) interactions may also play a role in SVO, as they do in other correlated \mbox{metals} where experiments~\cite{lanzara_evidence_2001, gadermaier_electron_phonon_2010, gerber_femtosecond_2017} and theory~\cite{yin_correlation-enhanced_2013, mandal_strong_2014, li_electron-phonon_2019} have highlighted the importance of $e$-ph coupling for spectral kinks~\cite{lanzara_evidence_2001, li_unmasking_2021} and electronic transport~\cite{abramovitch_combining_2023}. 
\mbox{A quantitative study} combining $e$-$e$ and $e$-ph interactions in SVO is needed to clarify the microscopic origin of its electronic behavior. 
\\
\indent
In this Letter, we show calculations of spectral and transport properties in SVO combining first-principles \mbox{$e$-ph} interactions with DFT+DMFT $e$-$e$ interactions~\cite{abramovitch_combining_2023}. 
We find that $e$-ph interactions govern the resistivity and its temperature dependence above $\sim$30~K, 
and account for the experimentally observed kinks 
and for most of the linewidth broadening of the spectral functions. 
In contrast, the $e$-$e$ interactions control the resistivity below 20 K, and are responsible for the most of the quasiparticle mass renormalization. 
We also find that the $e$-$e$ interactions lead to an enhancement of the effective
$e$-ph coupling. 
Our results provide a blueprint for quantifying electronic and lattice contributions to the properties of correlated metals.
\\
\indent
\textit{Electronic Structure and Electron-Phonon Coupling.\textemdash} 
We calculate the electronic structure, phonon dispersions, and $e$-ph coupling using DFT and density functional perturbation theory (DFPT) with the {\sc Quantum Espresso} package~\cite{Giannozzi_Quantum_2009, garrity_pseudopotentials_2014,zhou_ab_2021, floris_phonons_2022}. We use the experimental lattice parameter of 3.842 \AA~\cite{lan_structure_2003, ahn_low-energy_2022} and project the electronic structure onto the t$_{2g}$ $d$-orbitals of vanadium~\cite{arash_updated_wannier90_2014}. 
We use {\sc Perturbo} to compute the $e$-ph interactions, $e$-ph self-energy, spectral functions, and transport~\cite{zhou_perturbo_2021}. 
The $e$-$e$ self-energy is obtained with DFT+DMFT using the TRIQS code with a continuous-time quantum Monte Carlo solver~\cite{gull_ctmc_2011, parcollet_triqs_2015, seth_triqscthyb_2016, aichhorn_triqsdfttools_2016, Blaha_wien2k_2020, Beck_charge_2022} and Padé analytical continuation~\cite{parcollet_triqs_2015}. We use Hubbard-Kanamori interactions with $U = 4.5$ eV and $J = 0.675$ eV to produce spectra in agreement with experiment~\cite{yoshida_direct_2005,kobayashi_origin_2015, yoshida_correlated_2016}.
Additional computational details are provided in the Supplemental Material (SM)~\cite{supplemental_material}.
\\
\indent
As shown in Fig.~\ref{fig:bands-spec-imsigma}(a), our DFT calculations predict a bandwidth of 2.5 eV for the t$_{2g}$ electronic bands, which is renormalized by a factor $Z\! \approx \!0.5$ by DMFT, in agreement with experiments~\cite{yoshida_direct_2005} and previous DMFT results~\cite{nekrasov_momentum-resolved_2006, taranto_comparing_2013}. In the temperature range we study ($\sim$115$-$390~K), the imaginary part of the $e$-$e$ self-energy, $\mathrm{Im}\Sigma^{\rm e-e}$, follows a Fermi liquid behavior.  
Figure~\ref{fig:bands-spec-imsigma}(b) shows that $\mathrm{Im}\Sigma^{\rm e-e}(\omega,T)$ within 100 meV of the Fermi energy can be fit closely by a Fermi liquid parameterization~\cite{coleman2015introduction}, 
$\mathrm{Im}\Sigma^{\rm e-e}(\omega,T) = -c((\hbar\omega)^2 + \pi^2 (k_B T)^2)$ 
with $c \approx 0.33~\mathrm{eV}^{-1}$\footnote{In the following, we set $\hbar=k_B=1$}. 
Therefore, based on the Kramers-Kronig relations,  $\mathrm{Re}\Sigma^{\rm e-e}(\omega,T)$ and the quasiparticle dispersion near the Fermi energy depend weakly on temperature. 
\\
\indent
Figure~\ref{fig:bands-spec-imsigma}(c) shows the phonon dispersions in SVO computed with DFPT and color-coded according to the $e$-ph coupling strength $|g_\nu(\mathbf{q})|$, for each phonon mode $\nu$ and momentum $\mathbf{q}$, averaged over the Fermi surface (see SM~\cite{supplemental_material}). 
The $e$-ph coupling is stronger for the six highest-energy modes, 
which involve distortions of the VO$_6$ octahedra, such as Jahn-Teller modes.
\\
\indent
\textit{Spectral Properties.\textemdash}
We investigate the contributions of $e$-ph and $e$-$e$ interactions in SVO by computing the corresponding self-energies~\cite{abramovitch_combining_2023}. 
The real and imaginary parts of the $e$-$e$ and $e$-ph self-energies at 115 K are shown in Fig.~\ref{fig:se-spec}(a)-(b), respectively. 
The $e$-$e$ interactions dominate quasiparticle renormalization, as seen from the greater derivative of $\mathrm{Re}\Sigma^{\rm e-e}$ compared to $\mathrm{Re}\Sigma^{\rm e-ph}$ within 150~meV of the Fermi energy. 
Accordingly, extracting quasiparticle weights $Z = (1 - \frac{\partial \mathrm{Re}\Sigma(\omega)}{\partial \omega}|_{\omega = 0})^{-1}$ with a fit near the Fermi surface, gives a weak contribution to renomalization for $e$-ph interactions ($Z_{\rm e-ph} = 0.81$) and a dominant contribution for $e$-$e$ interactions, with $Z_{\rm e-e} = 0.53$ and $Z_{\mathrm{both}} = 0.47$. 
\\
\indent
The imaginary part of the self-energy shows an opposite behavior: $\mathrm{Im}\Sigma^{\rm e-ph}$ is much greater than $\mathrm{Im}\Sigma^{\rm e-e}$, and thus the $e$-ph interactions account for the majority of electron scattering and spectral width at low energy. 
The dominant role of $e$-$e$ interactions on quasiparticle renormalization in SVO, despite their small effect on low energy scattering consistent with a Fermi liquid, can be rationalized using the Kramers-Kronig relations~\cite{coleman2015introduction, hartnoll_planckian_2022}: 
due to the larger energy scales involved, the $e$-$e$ interactions dominate the imaginary part of the self-energy at higher energies (see SM~\cite{supplemental_material}), leading to a greater magnitude (and energy derivative) of $\mathrm{Re}\Sigma^{\rm e-e}(\omega)$ compared to $\mathrm{Re}\Sigma^{\rm e-ph}(\omega)$ at low energy.
\\
\indent
We compute the spectral function $A_{n\mathbf{k}}(\omega) = -\frac{1}{\pi}\mathrm{Im}G_{n\mathbf{k}}(\omega)$ from the Green's function 
\begin{equation}
    G_{n\mathbf{k}}(\omega,T) = \left[\omega - \varepsilon_{n\mathbf{k}} + \mu - \Sigma_{n\mathbf{k}}(\omega,T) \right]^{-1}
\end{equation}
at energy $\omega$ for electron band $n$ and momentum  $\mathbf{k}$. Here, $\varepsilon_{n\mathbf{k}}$ is the DFT band energy, $\mu$ is the Fermi energy, and $\Sigma_{n\mathbf{k}}(\omega,T)$ is the electron self-energy. Following our previous work~\cite{abramovitch_combining_2023}, in separate calculations we compute this Green's function using the self-energy from DMFT $e$-$e$ interactions, the lowest-order self-energy from $e$-ph interactions, and their sum~\footnote{The $e$-ph self-energy calculation uses the DFT electron propagator. Unlike in Sr$_2$RuO$_4$~\cite{abramovitch_combining_2023}, using the DMFT electron propagator has a negligible effect in SVO.}, obtaining corresponding spectral functions capturing different combinations of interactions [Fig.~\ref{fig:se-spec}(c)-(e)]. 
The spectral functions from $e$-ph, and those from $e$-ph plus $e$-$e$ interactions,  show a kink around 60 meV from the Fermi energy which has been observed in experiments~\cite{aizaki_self-energy_2012, yoshida_correlated_2016}. There is a corresponding sharp change in the derivative of $\mathrm{Re}\Sigma^{\rm e-ph}_{n\mathbf{k}}(\omega)$ at this energy [Fig.~\ref{fig:se-spec}(a)], whereas this feature is absent in $\mathrm{Re}\Sigma^{\rm e-e}_{n\mathbf{k}}(\omega)$. This result shows that the 60 meV kink observed experimentally in SVO is caused by $e$-ph interactions.
\\
\indent
\begin{figure*}[t!]
    \centering
    \includegraphics[width = \textwidth]{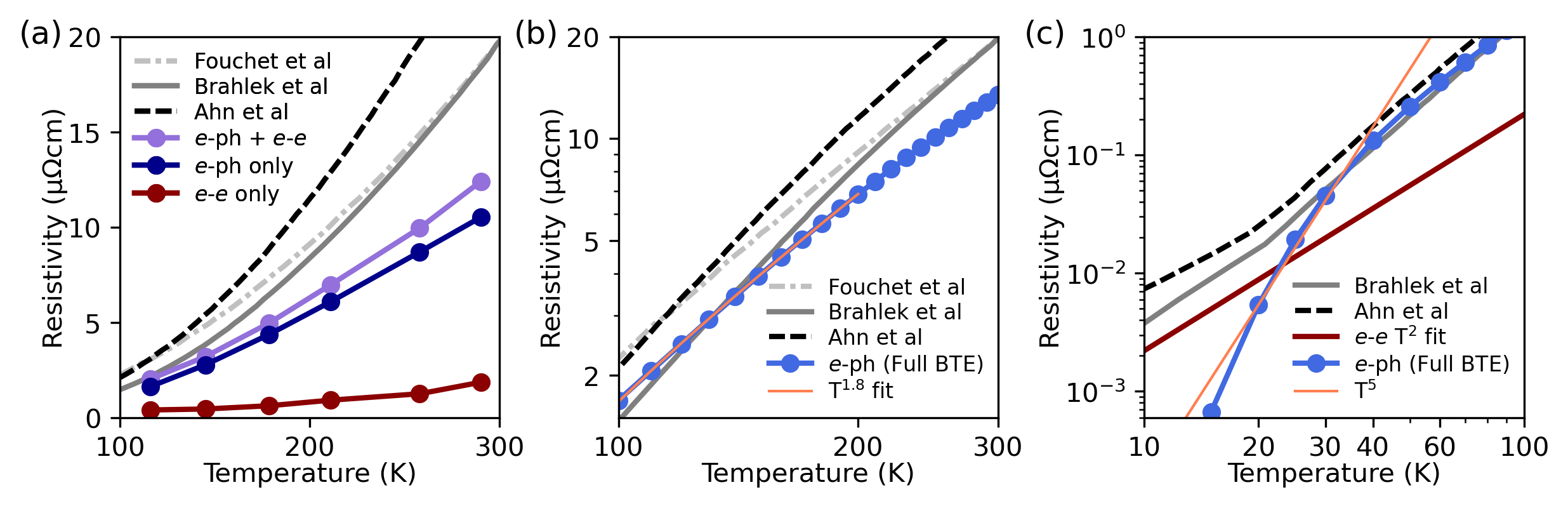}
    \caption{(a) Resistivity as a function of temperature calculated using the Green-Kubo formalism with $e$-$e$ interactions, \mbox{$e$-ph} interactions, and their combination. 
    (b) Temperature dependence of the $e$-ph limited resistivity calculated using the full (iterative) solution of the BTE. (c) Low-temperature $e$-ph and $e$-$e$ limited transport. 
    Experimental data (from which the $T=0$ residual resistivity was subtracted)    
    are from Refs.~\cite{fouchet_study_2016, ahn_low-energy_2022}. 
    }
    \label{fig:transport}
\end{figure*}
\textit{Transport.\textemdash}
Numerous experiments have measured a near $T^2$ temperature dependence of the resistivity in SVO below 300 K~\cite{inoue_bandwidth_1998, ardila_preparation_2000, lan_structure_2003, brahlek_accessing_2015, fouchet_study_2016,mirjolet_independent_2019,mirjolet_high_2019,berry_laser_2022, ahn_low-energy_2022, brahlek_hidden_2024}. 
Due to the strong electronic correlations in SVO, several studies have attributed this resistivity to $e$-$e$ interactions in the Fermi liquid regime~\cite{inoue_bandwidth_1998, fouchet_study_2016, berry_laser_2022}, where $T^2$ behavior is expected. An exception is recent work by Mirjolet \textit{et al.}, who argued that the temperature dependence is better explained by $e$-ph limited resistivity with strong coupling to a dominant phonon mode~\cite{mirjolet_electronphonon_2021}. 
Recently, the growth of ultra-clean samples has enabled detailed measurements of the resistivity with reduced defect scattering~\cite{brahlek_accessing_2015, ahn_low-energy_2022, brahlek_hidden_2024}. In these samples, Ahn \textit{et al.}~\cite{ahn_low-energy_2022} and Brahlek \textit{et al.}~\cite{brahlek_hidden_2024} find a near-$T^2$ resistivity below 25 K and between about 100$-$300~K, together with a stronger than $T^2$ temperature dependence at intermediate temperatures.
\\ 
\indent 
To understand the microscopic origin of this behavior, we compute the resistivity arising from $e$-ph and $e$-$e$ interactions using the Green-Kubo formula~\cite{mahanManyParticle2000, abramovitch_combining_2023}: 
\vspace{-5pt}
\begin{multline}
    \!\!\!\!\!\!\rho^{-1}_{\alpha\beta}(T) = \frac{\pi \hbar e^2}{V_{uc}} \int\! d\omega  (-f'(\omega,T)) \sum_{nk}v^\alpha_{n\mathbf{k}}v^\beta_{n\mathbf{k}}A_{n\mathbf{k}}(\omega,T)^2,
\label{eq:greenkubo}
\end{multline}
where $\rho_{\alpha\beta}$ is the resistivity tensor, $\alpha$ and $\beta$ are Cartesian directions, $f'(\omega,T)$ is the energy derivative of the Fermi occupation factor, $v_{n\mathbf{k}}^\alpha$ is the band velocity, and $A_{n\mathbf{k}}(\omega,T)$ is the spectral function. 
The resistivity for different combinations of interactions is shown in Fig.~\ref{fig:transport}(a) and compared with experimental data~\cite{fouchet_study_2016, ahn_low-energy_2022, brahlek_hidden_2024}. 
\\
\indent
Surprisingly, we find that the resistivity is governed by the $e$-ph interactions in SVO. The $e$-ph limited resistivity is an order of magnitude greater than the $e$-$e$ limited resistivity, with the latter accounting for only $\sim$10\% of the experimental value. This result is in contrast with the conventional wisdom that transport properties in SVO are governed by purely electronic interactions. In addition, the contributions are opposite to another prototypical strongly correlated metal, $\rm{Sr_{2}RuO_{4}}$, where the $e$-ph interactions account for only $\sim$10\% of the resistivity~\cite{abramovitch_combining_2023}. 
In SVO, the $e$-ph contribution is similar in magnitude to the experimental value, and the total resistivity including both interactions is in good agreement with experiments. 
Below, we show that taking into account the electron correlation induced enhancement of the 
$e$-ph interactions increases the resistivity and brings the results in even better agreement with experiments.  
\\
\indent
The temperature dependence of the $e$-ph limited resistivity is analyzed in more detail in Fig.~\ref{fig:transport}(b), where we show our results on a log-log plot and compare them with experiments~\cite{fouchet_study_2016, ahn_low-energy_2022, brahlek_hidden_2024}. 
In that plot, the resistivity is computed with the full (iterative) solution of the Boltzmann transport equation (BTE)~\cite{zhou_perturbo_2021} to include backscattering and improve the treatment of acoustic phonons. 
The computed $e$-ph limited resistivity follows a $T^{1.8}$ temperature dependence between 100$-$200 K, in excellent agreement with the $T^{1.8-2}$ dependence found in experiments in that temperature range~\cite{fouchet_study_2016,ahn_low-energy_2022, brahlek_hidden_2024}, and falls to $T^{1.5}$ at 300 K. We identify the origin of this nearly $T^2$ temperature trend of the $e$-ph limited resistivity by analyzing the contribution of different phonon modes. Our calculations show that the increasing contribution of strongly coupled optical phonons at higher temperatures is responsible for the $T^{2}$ dependence of the resistivity between 100$-$200 K (see SM~\cite{supplemental_material}). 
\\
\indent
Next, we focus on transport at low temperature, where the $e$-ph contribution is expected to be weaker. While DMFT calculations become difficult at low temperatures, we obtain the $e$-$e$ limited resistivity by extrapolating our higher-temperature DMFT calculations with a $T^2$ fit. This approach is justified because the $e$-$e$ scattering is in the Fermi liquid regime below at least 400 K. 
Figure~\ref{fig:transport}(c) shows the computed $e$-ph and DMFT $e$-$e$ limited resistivities below 100~K. 
We find a clear crossover between 20$-$30 K from $e$-ph to $e$-$e$ dominated transport. The $e$-ph limited resistivity becomes much smaller than the $e$-$e$ limited resistivity below 20 K, showing that $e$-$e$ scattering governs transport at low temperature. This result indicates that $e$-$e$ interactions are the origin of the $T^2$ resistivity observed experimentally below 25 K, with the $e$-ph contribution causing deviations from a $T^2$ behavior above 25 K. Note that our DMFT resistivity underestimates the experimental value below 25 K  by a factor of 2-3. We attribute this discrepancy to limitations in the single site DMFT impurity model, as transport can be sensitive to the methods used to downfold orbitals and effective interactions~\cite{deng_transport_2016,petocchi_screeening_2020, hampel_correlation_2021, abramovitch_combining_2023}, the lack of non-local contributions to the DMFT self-energy~\cite{nomura_nonlocal_2015}, and corrections to the current-current vertex~\cite{haule_dynamical_2010, vucicevic_conductivity_2019, mu_optical_2022, mu_adequacy_2024}. 
\\
\indent
\textit{Correlation-corrected electron-phonon interactions.\textemdash} 
Strong electronic interactions are known to significantly modify $e$-ph interactions~\cite{kulic_influence_1994, huang_electron_phonon_2003, li_competing_2017, scazzola_competing_2023}. In correlated metals, $e$-ph coupling is often enhanced. 
For example, calculations using hybrid functionals and the $GW$ method found correlation-enhanced $e$-ph coupling in unconventional superconductors, attributing the enhancement to decreased electronic screening~\cite{yin_correlation-enhanced_2013}. Similarly, in multiband $d$-electron systems such as FeSe, treating correlations with DMFT enhances $e$-ph coupling, in this case by increasing the orbital polarization response to phonon perturbations~\cite{mandal_strong_2014}. 
\\
\indent
To study the role of correlations in SVO, we compute the $e$-ph interactions using Hubbard-corrected DFPT (DFPT+$U$)~\cite{floris_phonons_2022, zhou_ab_2021}. DFT+$U$ can be viewed as a static approximation of DFT+DMFT~\cite{park_computing_2014}; DFT+DMFT $e$-ph coupling calculations are not currently possible. Like DFT+DMFT, DFT+$U$ captures the strong local interactions between $d$-orbitals and the resulting change in orbital polarization response, which modifies $e$-ph coupling. We use a Hubbard-$U$ parameter of 3 eV in the atomic orbital basis, which is somewhat empirical but provides orbital polarization and band structure responses to phonon perturbations similar to a calculation including a Hubbard-$U$ parameter from linear response and a reasonable value of Hund's $J$. The orbital occupation responses are also similar to those found using DMFT with the same orbitals and interaction parameters used throughout (see SM~\cite{supplemental_material})
\\
\indent
Adding the Hubbard correction has a small effect on the phonon dispersions, but it enhances the $e$-ph interactions, as shown in Fig.~\ref{fig:hub_epc_res}(a). The enhancement is mode-dependent and is generally higher for strongly-coupled optical phonons involving VO$_6$ distortions. 
The enhancement is also higher for phonons with momenta away from the $\Gamma$ point, suggesting a more important role of correlations for distortions breaking lattice-translation symmetry. The spectral and transport properties computed with enhanced $e$-ph coupling from DFPT+$U$ give results qualitatively similar to those from DFPT~\cite{supplemental_material}, but with stronger $e$-ph effects. 
Notably, the $e$-ph limited resistivity increases by $\sim$35\% at room temperature [Fig.~\ref{fig:hub_epc_res}(b)], bringing the resistivity computed with both $e$-ph and $e$-$e$ interactions into very good agreement with experiments. For example, the computed resistivity at 290~K is 16~$\mu \Omega$cm, versus an experimental value of 19$-$25~$\mu \Omega$cm~\cite{fouchet_study_2016, ahn_low-energy_2022, brahlek_hidden_2024}.  
\begin{figure}
    \centering
    \includegraphics[width = 0.5\textwidth]{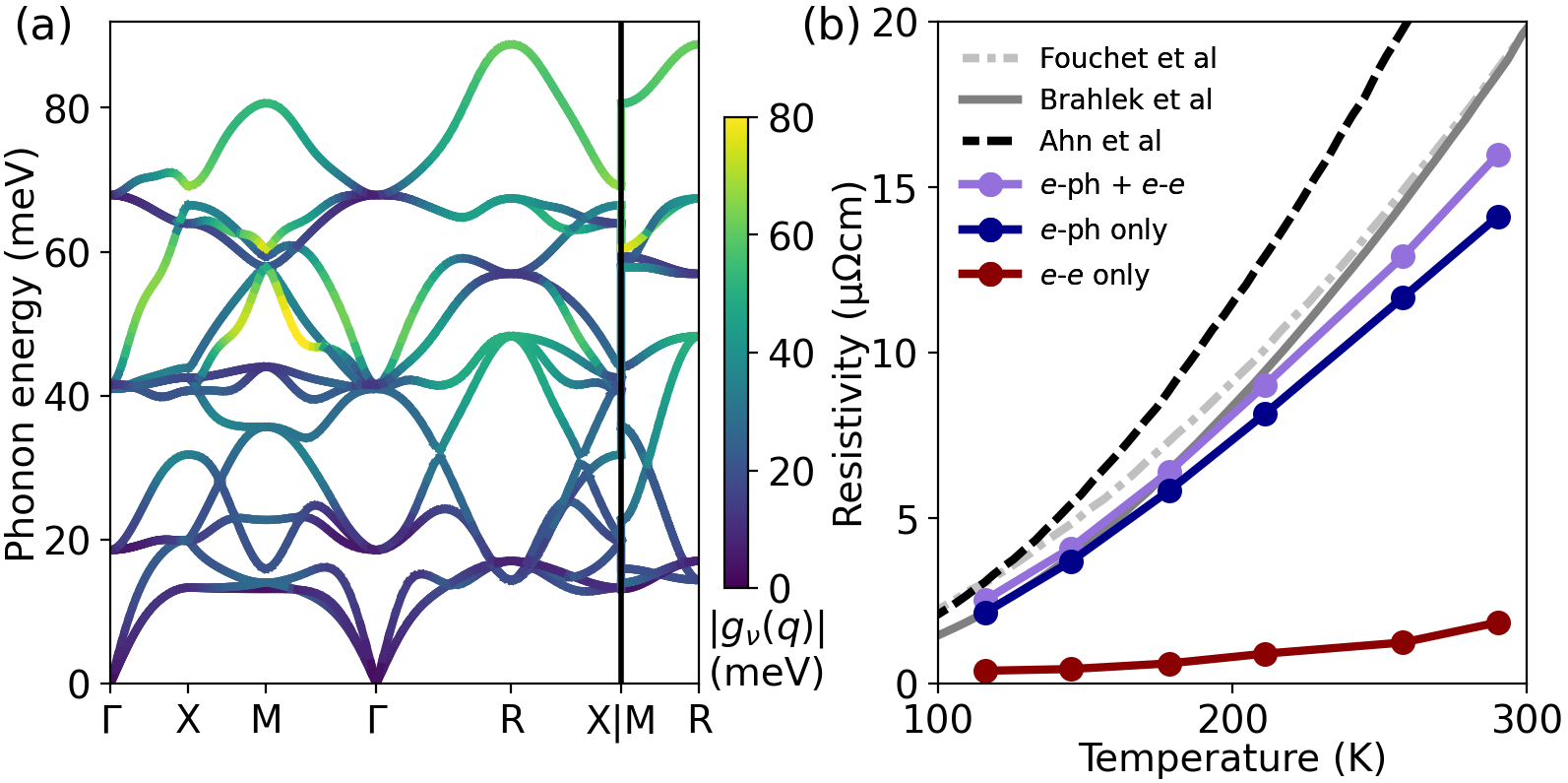}
    \caption{(a) Phonon dispersions and $e$-ph coupling as in Fig.~\ref{fig:bands-spec-imsigma}(c) but calculated with DFPT+$U$. Note the change in $e$-ph coupling scale. (b) Transport as in Fig.~\ref{fig:transport}(a) but calculated with DFPT+$U$ phonons and $e$-ph couplings, showing improved agreement with experiments~\cite{fouchet_study_2016, ahn_low-energy_2022, brahlek_hidden_2024}.}
    \label{fig:hub_epc_res}
\end{figure}
\\
\indent
\textit{Discussion.\textemdash} 
The origin of the temperature dependence of the resistivity merits further discussion. While the $T^2$ trend for $e$-$e$ interactions is expected based on Fermi liquid theory~\cite{mahanManyParticle2000}, the origin of the near-$T^2$ behavior of the $e$-ph limited resistivity is less clear. 
At very low temperatures, the $e$-ph limited resistivity in metals is expected to exhibit a $T^5$ temperature dependence~\cite{bass_temperature_1990} when scattering is dominated by acoustic phonons with momentum $q \propto k_BT$. 
In our calculations, we find a $T^5$ $e$-ph limited resistivity below $\sim$30 K, but the overall resistivity becomes $e$-$e$ limited in this temperature range, explaining the experimental $T^2$ resistivity below 25 K. 
\\ 
\indent
In the high-temperature limit, based on the temperature dependence of the phonon occupations, one expects a $T$-linear $e$-ph limited resistivity~\cite{bass_temperature_1990}. 
However, this requires that all phonon modes contribute equally to $e$-ph scattering, with no mode frozen out. While our computed $e$-ph limited resistivity becomes nearly $T$-linear well above 300~K, it is close to a $T^2$ behavior between $\sim$100$-$200~K, in agreement with experiments. 
As discussed above, this $T^2$ trend is due to the increasing contribution of higher-energy optical phonons with strong $e$-ph coupling for increasing temperatures~\cite{supplemental_material}. 
Note also that $e$-ph scattering above $\sim$50~K in SVO involves phonons with all momenta, ruling out momentum-dependent mechanisms resulting in $T^2$ behavior~\cite{kukkonen_t_1978}. 
\\
\indent
Finally, we analyze two approximations made in the $e$-ph transport calculations (see results in SM~\cite{supplemental_material}). First, we examine the use of the lowest-order $e$-ph self-energy~\cite{migdal_interaction_1958, prange_transport_1964} by computing the resistivity with a cumulant diagram-resummation method capable of treating delocalized polarons~\cite{zhou_predicting_2019}. 
Including polaron effects leads to a small increase in the resistivity, showing that lowest-order $e$-ph interactions are adequate to describe SVO. Second, we examine the effect of vertex corrections to the current-current correlation function in the Green-Kubo formalism~\cite{mahanManyParticle2000,coleman2015introduction}. 
Vertex corrections improve the description of backscattering and the momentum dependence of $e$-ph scattering, which is particularly important at low temperature~\cite{mahanManyParticle2000}. 
We assess their role above 100 K in the semiclassical limit by comparing the full solution of the BTE, which includes vertex corrections, to the relaxation time approximation, which neglects them. 
We find that vertex corrections in the BTE give only a small increase in the resistivity and its temperature dependence. This analysis shows that higher-order $e$-ph interactions and vertex corrections play a minor role in SVO and do not affect our conclusions. 
\\
\indent
\textit{Conclusion.\textemdash}
In summary, we have shown that $e$-ph interactions play an essential role in
the transport and spectral properties of a prototypical correlated metal, SVO. 
In this material, electronic correlations control 
other aspects of the low-energy physics, 
including the quasiparticle mass renormalization and transport at low temperature. We also found that electronic correlations lead to an effective enhancement of the $e$-ph interactions. 
This suggests that SVO may serve as a testbed for investigating the interplay between electron correlations and $e$-ph interactions. 
Our results highlight the potential of first-principles calculations combining $e$-$e$ and $e$-ph interactions in a consistent way as an emerging tool to study correlated materials. 
This work paves the way for a quantitative description of transport and spectral properties 
in broad classes of correlated quantum \mbox{materials.}\\ 

\textit{Acknowledgements.\textemdash} 
We thank Andrew Millis, Jennifer Coulter, and Roman Engel-Herbert for helpful discussions. 
D.J.A. is supported by the National Science Foundation Graduate Research Fellowship under Grant No. 2139433. 
This work was also supported by the National Science Foundation under Grant No. OAC-2209262, which provided for code development. 
D.J.A. and M.B. were partially supported by the AFOSR and Clarkson Aerospace under Grant No. FA95502110460. 
J.-J.Z. acknowledges support from the National Key R\&D Program of China (Grant No. 2022YFA1403400). J.M. is supported by the Slovenian Research Agency (ARIS) under Grants No. P1-0044 and J1-2458. 
This research used resources of the National Energy Research
Scientific Computing Center, a DOE Office of Science User Facility
supported by the Office of Science of the U.S. Department of Energy
under Contract No. DE-AC02-05CH11231 using NERSC award
NERSC DDR-ERCAP0026831. The Flatiron Institute is a division of the Simons Foundation. 
\bibliography{apssamp}

\end{document}